\documentclass[aps,prb,twocolumn, superscriptaddress, showpacs]{revtex4-1}
\usepackage{amsmath}
\usepackage{amssymb,amsfonts}
\usepackage{graphicx}
\usepackage{mathtools}
\usepackage{relsize}
\usepackage{lmodern}
\usepackage{slantsc}
\usepackage{scalefnt}
\usepackage{subfigure}
\usepackage{dsfont}
\usepackage{xspace} 
\usepackage{ifpdf}
\usepackage{array}
\usepackage{color}
\allowdisplaybreaks[1]

\newcommand{\be}{\begin{equation}}
\newcommand{\ee}{\end{equation}}
\newcommand{\bse}{\begin{subequations}}
\newcommand{\ese}{\end{subequations}}

\newcommand{\Z}{\mathbb{Z}}
\newcommand{\ii}{\mathrm{i}}

\newcommand{\T}{\mathcal{T}}
\newcommand{\A}{\mathcal{A}}

\newcommand{\bpm}{\begin{pmatrix}}
\newcommand{\epm}{\end{pmatrix}}
\newcommand{\bmm}{\begin{matrix}}
\newcommand{\emm}{\end{matrix}}

\newcommand{\x}{\times}

\newcommand{\FF}{$\text{Fibo}\times\overline{\text{Fibo}}$\xspace}

\makeatletter
\newcommand*{\Relbarfill@}{\arrowfill@\Relbar\Relbar\Relbar}
\newcommand*{\xeq}[2][]{\ext@arrow 0055\Relbarfill@{#1}{#2}}
\makeatother

\usepackage{varioref}

\begin{document}

\title{Ground State Degeneracy of Topological Phases on Open Surfaces}

\author{Ling-Yan Hung}
\email{jhung@perimeterinstitute.ca}
\affiliation{Department of Physics and Center for Field Theory and Particle Physics, Fudan University, Shanghai 200433, China}
\affiliation{Department of Physics, Harvard University, Cambridge MA 02138, USA}
\author{Yidun Wan}
\email{ywan@perimeterinstitute.ca}
\affiliation{Perimeter Institute for Theoretical Physics, Waterloo, ON N2L 2Y5, Canada}

%\date{August 28, 2014}

\begin{abstract}
We relate the ground state degeneracy (GSD) of a non-Abelian topological phase on a surface with boundaries to the anyon condensates that break the topological phase to a trivial phase. Specifically, we propose that gapped boundary conditions of the surface are in one-to-one correspondence to the sets of condensates, each being able to completely break the phase, and we substantiate this by examples. The GSD resulting from a particular boundary condition coincides with the number of confined topological sectors due to the corresponding condensation. These lead to a generalization of the Laughlin-Tao-Wu (LTW) charge-pumping argument for Abelian fractional quantum Hall states (FQHS) to encompass non-Abelian topological phases, in the sense that an anyon loop of a confined anyon winding a non-trivial cycle can pump a condensed anyon from one boundary to another. Such generalized pumping may find applications in quantum control of anyons, eventually realizing topological quantum computation.
\end{abstract}
\maketitle
%\tableofcontents
%\makeatletter
%\let\toc@pre\relax
%\let\toc@post\relax
%\makeatother

\section{Introduction}\label{sec:intro}

A key feature of intrinsic topological orders is the existence of protected
GSD. On closed spatial 2-surfaces, its genus number and the fusion rules between anyon excitations determine the GSD\cite{TaoWu1984,Niu1985,Wen1989,Wen1990,Wen1990c,Wen1991,Wen1990a, Nayak2008}.
Protected GSD is vital to topological quantum computation, and yet realizing GSD on high genus closed surfaces is unfeasible in experiments.  Obviously it is much more natural to build finite open systems. Yet, it is necessary that any boundary massless modes that often appear can be gapped to have a well defined GSD. The gapping
conditions of Abelian phases have recently been understood in terms of the concept of Lagrangian subsets\cite{Kitaev2012,Levin2013,Barkeshli,Barkeshli2013c,Wang2012,HungWan2013a}, and subsequently the GSDs of these Abelian phases on open surfaces with multiple boundaries were computed\cite{Wang2012,Iadecola2014}, based on the idea of anyon transport across boundaries. Experiments detecting and applying the topological degeneracy with gapped boundaries were proposed in \cite{Barkeshli2014,Barkeshli2014a}. Nevertheless, non-Abelian phases bear much richer sets of degenerate ground states, and the braiding of non-Abelian anyons serves as the best known candidate that may realize universal topological quantum computing. It is also hopeful to realize or simulate non-Abelian anyons\cite{Lesanovsky2012}. Despite the importance of understanding non-Abelian phases on open surfaces, it remains a big open problem, a problem we shall solve here via the method of anyon condensation. 
Summarizing our main results: 
\begin{itemize}
\item We find the condition for gapped boundaries of non-chiral non-Abelian phases by identifying each such boundary to a set of anyon condensate; the condition allows one to classify all gapped boundaries for the given phase.
{\bf  Our results also encompass situations in which a defect/phase boundary  separates arbitrary phases because any such system can always be mapped back to one where a phase ends on the trivial vacuum by the folding trick.} 
\item for any given boundary conditions on some arbitrary open system, we describe the computation of the GSD, dictated by a reduced set of conserved topological sectors---the set of \emph{confined anyons}---anyons mutually non-local with the boundary condensate. Typically, on a cylinder,
\be
 \mathrm{GSD}=\# \text{confined anyons}.
 \ee
We show an explicit non-trivial example of a GSD counting on a cylinder whose two ends have distinct boundary  conditions.  {\bf Note that such reduction in the number of conserved anyons suggests a novel way to engineer desired conserved bulk anyons suitable for specific quantum computations;} Our method is corroborated using the correspondence between a generic non-Abelian phase $\A$ and its double $\A\x \bar\A$ via the folding trick.

\item we find connections between our counting and prior works on Abelian phases that make use of charge transport. This generalizes the LTW charge-pump argument in Abelian FQHS\cite{Laughlin1983,TaoWu1984} to generic anyon-pump in non-Abelian phases, hinting at novel ways of braiding and controlling non-Abelian anyons. Such anyon transport may also help experimentally discern different topological phases \cite{Barkeshli2014,Barkeshli2014a};

\end{itemize}
We shall make heavy use of the technologies studying anyon condensation\cite{Bais2002,Bais2009,Bais2009a}.
The basic premise of anyon condensation is that certain types of anyons cease to have conserved particle number across a phase transition; they thus effectively condense, exactly as how Cooper pairs condense, in the process breaking some symmetry. As in usual Bose condensation, the condensable anyons should have bosonic self-statistics. There are various constraints that determine the properties of the condensed phase, such as the types of anyons that remain conserved and their fusion rules.
We shall refer the reader to the original references for details. Some prior attempts on non-Abelian phases are found in \cite{Bombin2008,Bombin2011}.

\section{GSD of the $\Z_2$ toric code on open surfaces revisited}
\begin{figure}[ht]
\includegraphics[width=0.25\textwidth]{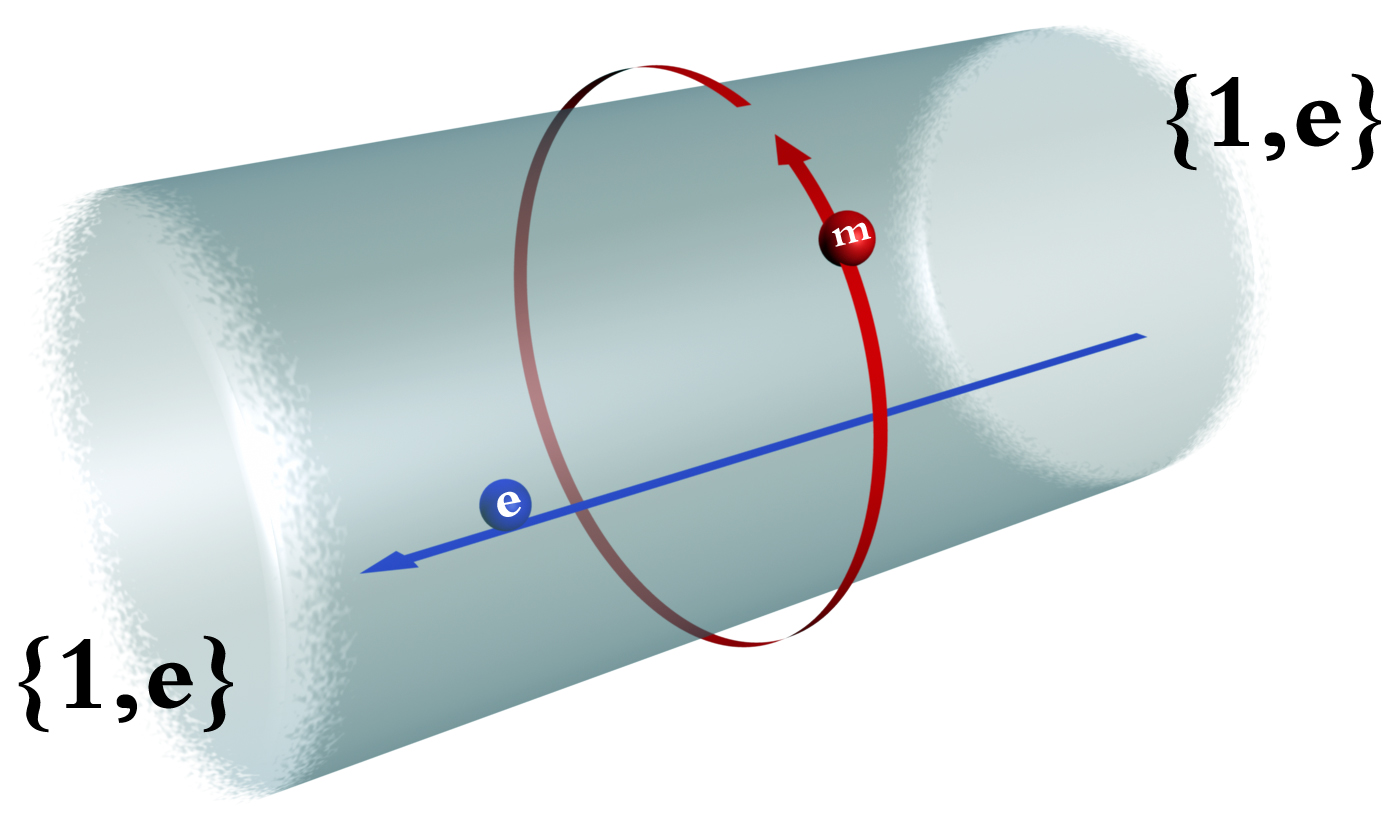}
\caption{Ground state basis specified by a conserved anyon line $m$ winding the cylinder where the boundaries are characterized by the $L_e$ condensate.
A unit of $e$ is transferred across the boundaries via the LTW charge pumping mechanism if the unit $m$ line can be changed to two units adiabatically.}
\label{fig:toricCyl}
\end{figure}

To explain the generalization of Lagrangian subsets and the subsequent GSD counting in Abelian phases, 
we revisit these concepts from the perspective of anyon condensation by taking the $\Z_2$ toric code as an example. Recall that the $\Z_2$ toric code has four topological sectors $\{1,e,m,f\}$, where $e,m$ are self-bosons, and $f$ a fermion. Any two distinct nontrivial anyons are mutual-fermions. The nontrivial fusion rules are $e^2=m^2=f^2=1$ and $e\x m=f$. 
Each boundary of the system admits two distinct gapping conditions, respectively characterized by the two
\emph{Lagrangian subsets}:
\be\label{eq:tcLagSets}
L_e = \{1,e\}, \,\,\, L_m = \{1,m\}.
\ee
A Lagrangian subset $L$ is a maximal collection of anyons that are self-bosons with trivial mutual statistics and that all remaining anyons excluded from the set are non-local wrt at least one member of $L$ \cite{Kitaev2012,Levin2013,Barkeshli,Barkeshli2013c}. This is clearly satisfied by both sets in \eqref{eq:tcLagSets}. 
One crucial observation\cite{Kitaev2012,Levin2013,Wang2012,Bais2002,Bais2009,Bais2009a,Hung2013,Gu2014a} is that for the boundary condition characterized by a set $L_i$, an anyon in the set ceases to be conserved and can be either created or annihilated at the boundary. Therefore, a gapped boundary condition $L_i$ is equivalent to the condensation of the anyons in $L_i$ right at the boundary. Any anyon not in a condensed $L_i$ would be \emph{confined} at the boundary\cite{Bais2002,Bais2009,Bais2009a,Hung2013,Gu2014a,Bais2014}. For example, $m$ and $f$    are confined in $L_e$ condensate and thus are mobile in the bulk but fail to cross the boundary into the vacuum. Equally importantly, in the vicinity of the $L_e$ condensate, $m$ and $f$ are indistinguishable, like $1$ and $e$ become identified, by fusion with any number of $e$'s freely supplied by the boundary condensate\cite{Bais2002,Bais2009,Bais2009a,Hung2013,Gu2014a,Bais2014}.  This leads to an easy GSD counting. Consider  a cylinder with both boundaries characterized by $L_e$. A convenient basis for ground states consists of uncontractible anyon loops winding the cylinder. For the $L_e$ boundaries, only two distinct anyon loops exist: $1$ and $m$. One then infers that 
\be
GSD^{L_e \textrm{boundaries on cylinder}}_{\Z_2 \textrm{toric code}} =2 ,
\ee 
in accord with the result of \cite{Wang2012}. This is illustrated in Fig. \ref{fig:toricCyl}.
The LTW charge-pump argument for FQHS\cite{Laughlin1983,TaoWu1984} applies here: If one threads a magnetic flux loop around the cylinder and adiabatically increase it from zero to  a unit $m$ flux, a charge $e$ can be pumped from one boundary to the other of the cylinder, as depicted in Fig. \ref{fig:toricCyl}.

To deal with more boundaries potentially characterized by different $L_i$'s, we need only to work out the remaining conserved (i.e. confined) distinct uncontractible anyon loops, and the anyon loops around different cycles must admit at least a fusion channel. We now generalize the procedure described above  to non-Abelian phases.

%%%%%%%%%%%%%%%%%%%%%%%%%%%%%%%%%%%%%%%%%%%%%%%%%%%%%%%%%

 %%%%%%%%%%%%%%%%%%%%%%%%%%%%%%%%%%%%%%%%%%%%%%%%
\section{GSD of Non-Abelian phases on open surfaces}\label{sec:genPrinciple}
We now lay down the general procedure to obtain the GSD
of a generic non-Abelian topological order with boundaries.  We will 
illustrate each step with the example of the doubled Fibonnacci model. To avoid clutter, 
the topological data of the doubled Fibonnacci model and notations are 
reviewed in the supplemental material.  

\textbf{Defining boundary conditions.}
First we have to decide upon the boundary condition on each boundary.
Each boundary whose edge modes could be completely gapped is characterized by
a generalized Lagrangian subset $L$, which is a collection of anyons that could condense
simultaneously at the boundary, and that the  resultant phase after the condensation $\T_L$
contains only confined anyons as well as the trivial sector. 
As reviewed in the supplemental material, in the case of \FF, it has $L_{\tau\bar\tau}=\{1,\tau\bar\tau\}$, leading to 
$\T_{L_{\tau\bar\tau}}= \{1,\chi\}$, where $\chi$ behaves like a Fibonacci anyon $\tau$ except for its lack of
a well-defined topological spin.

\textbf{Counting GSD via confined charges.}
If all the boundaries are characterized by the same $L$, the GSD is obtained
as follows. We first find out the fusion rules of all the confined anyons in $\T_L$. Then we count all possible
basis states constructed from loops of confined anyons winding nontrivial cycles. This is subjected
to the consistency condition on anyons wrapping cycles that merge have to fuse to the anyon wrapping
the resultant merged cycle. This is to ensure no net charge exists in the bulk. 

Consider for example \FF on a cylinder, where both boundaries must be characterized by $L_{\tau\bar\tau}$. As a result, the only conserved nontrivial topological sector must be the confined $\chi\in \T_{L_{\tau\bar\tau}}$, as it cannot leak through the boundaries into the vacuum.  We conclude that the 2 distinct sectors in $\T_L$ implies that
\be \label{eq:fiboGSDcyl}
GSD^{\text{cylinder}}_{\text \FF}=|\T_{L_{\tau\bar\tau}}|=2,
\ee   

\begin{figure}[ht!]
\centering
\subfigure[]{\includegraphics[width=0.2\textwidth]{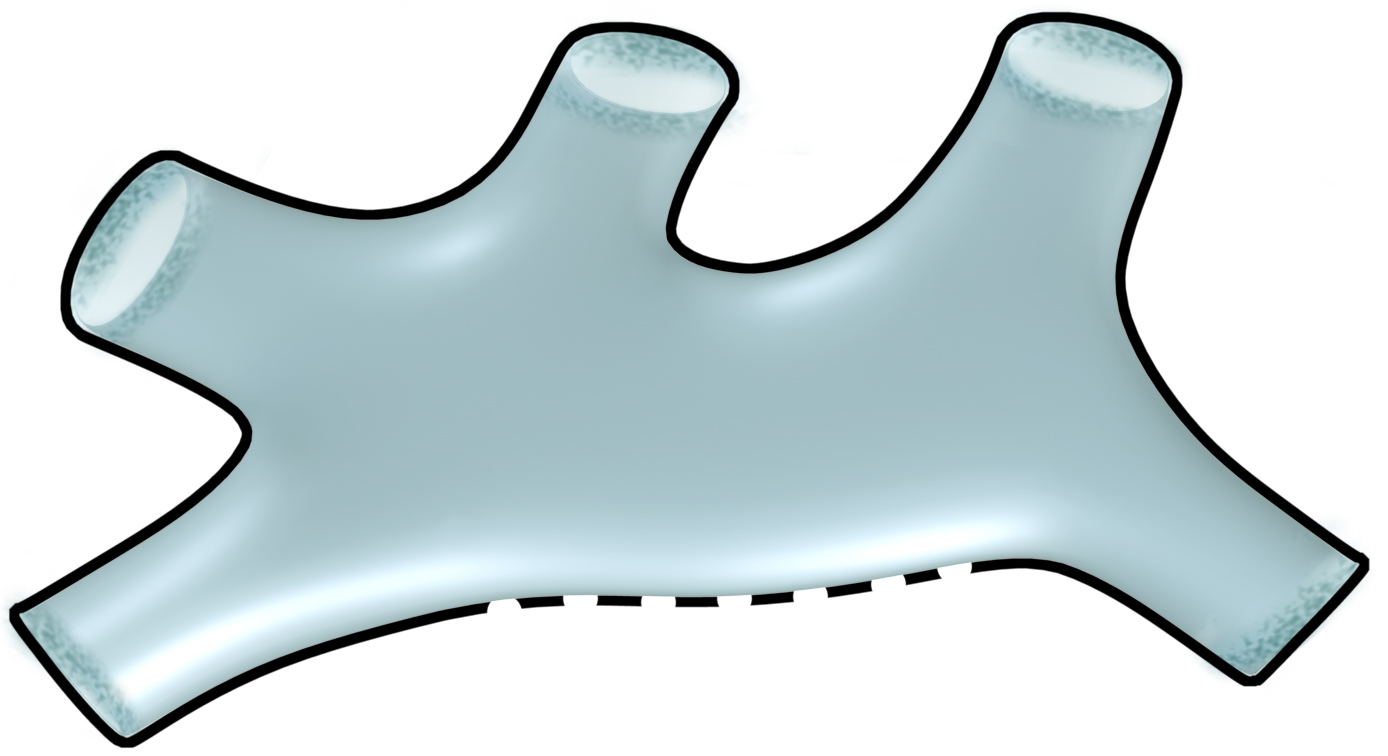}
\label{fig:multiHole}
}
\subfigure[]{\includegraphics[width=0.2\textwidth]{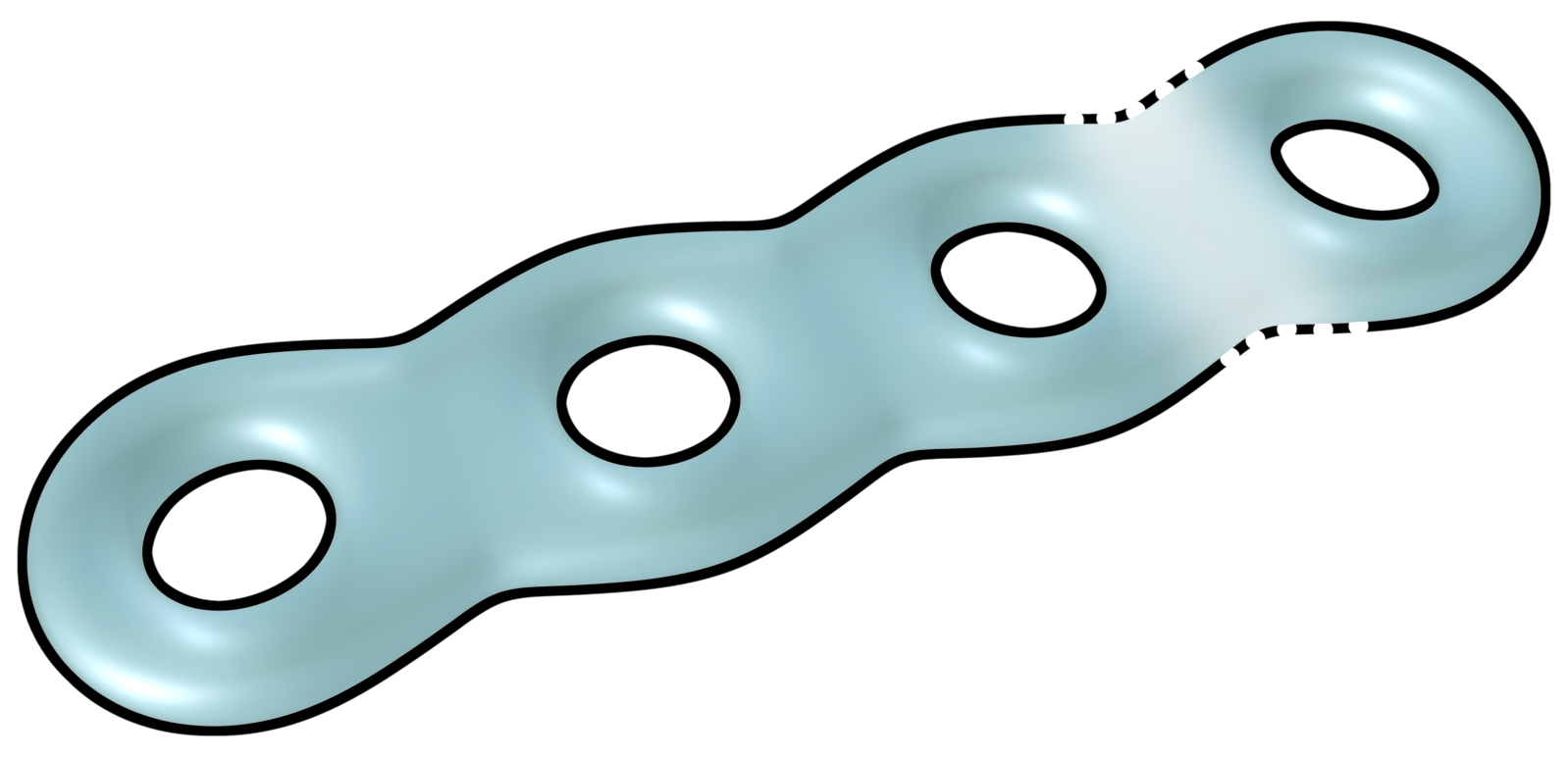}
}
\caption{(a) An open surface with $M>3$ boundaries. (b) A genus-$(M-1)$ torus. }\label{fig:multiHoleGenus}
\end{figure}

To check our claims, note that it is expected that
\be\label{eq:fiboID}
GSD^{\text{cylinder}}_{\text \FF}=GSD^{\text{torus}}_{\text{Fibo}}.
\ee
This is because \FF ending at a boundary can be thought of as folding up a Fibonacci phase characterized by single copy of the anyons $\{1,\tau\}$. Thus the \FF on a cylinder is in fact equivalent to the $\text{Fibonacci}$ phase itself residing on two different cylinders, yet joined at both boundaries because of the gapped boundary condition we imposed on \FF. That is, we have in fact $\text{Fibo}$ on a torus.
\begin{figure}[ht!]
\centering
\subfigure[]{\includegraphics[width=0.2\textwidth]{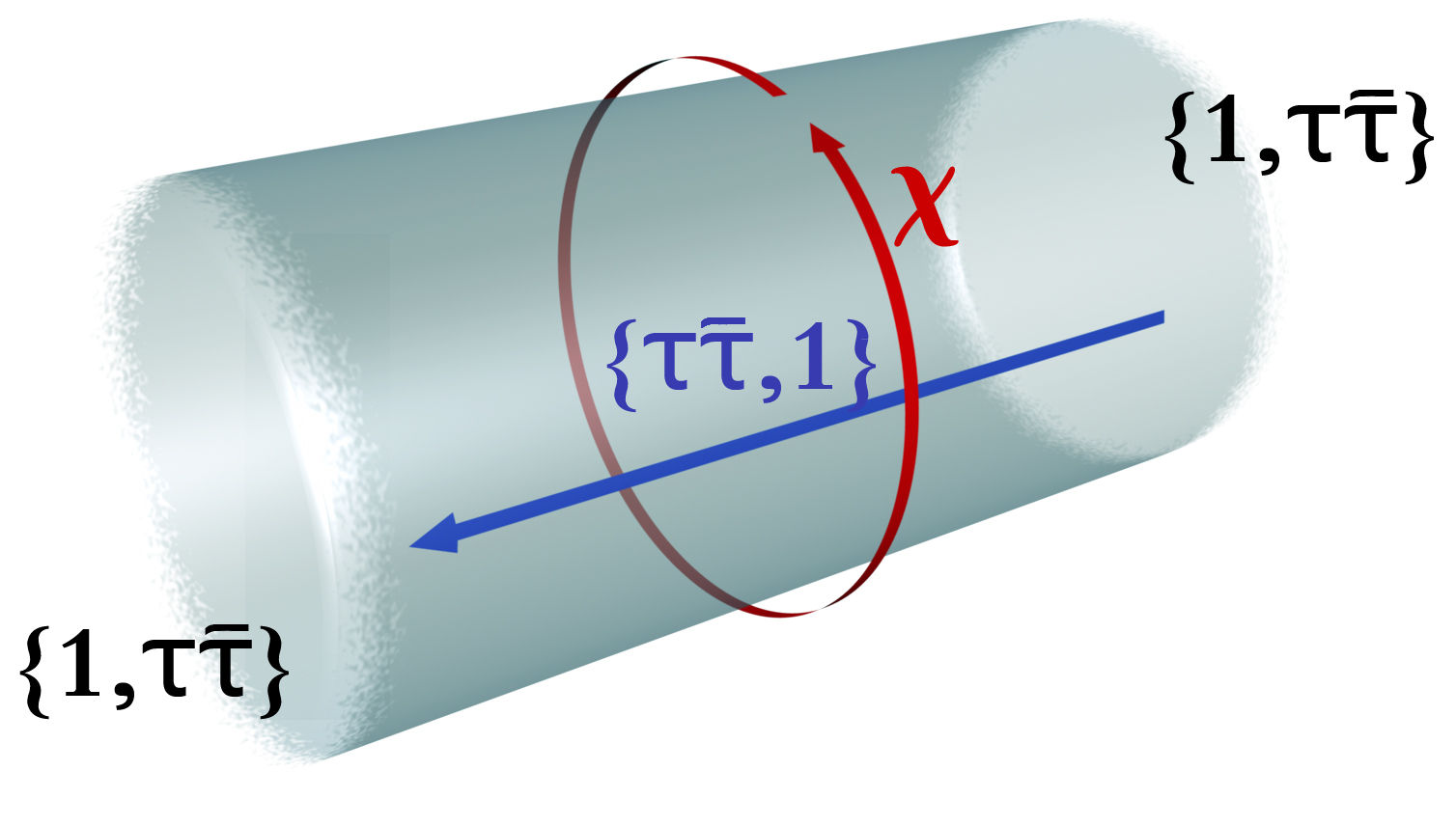}
\label{fig:fiboCylinder}
}
\subfigure[]{\includegraphics[width=0.15\textwidth]{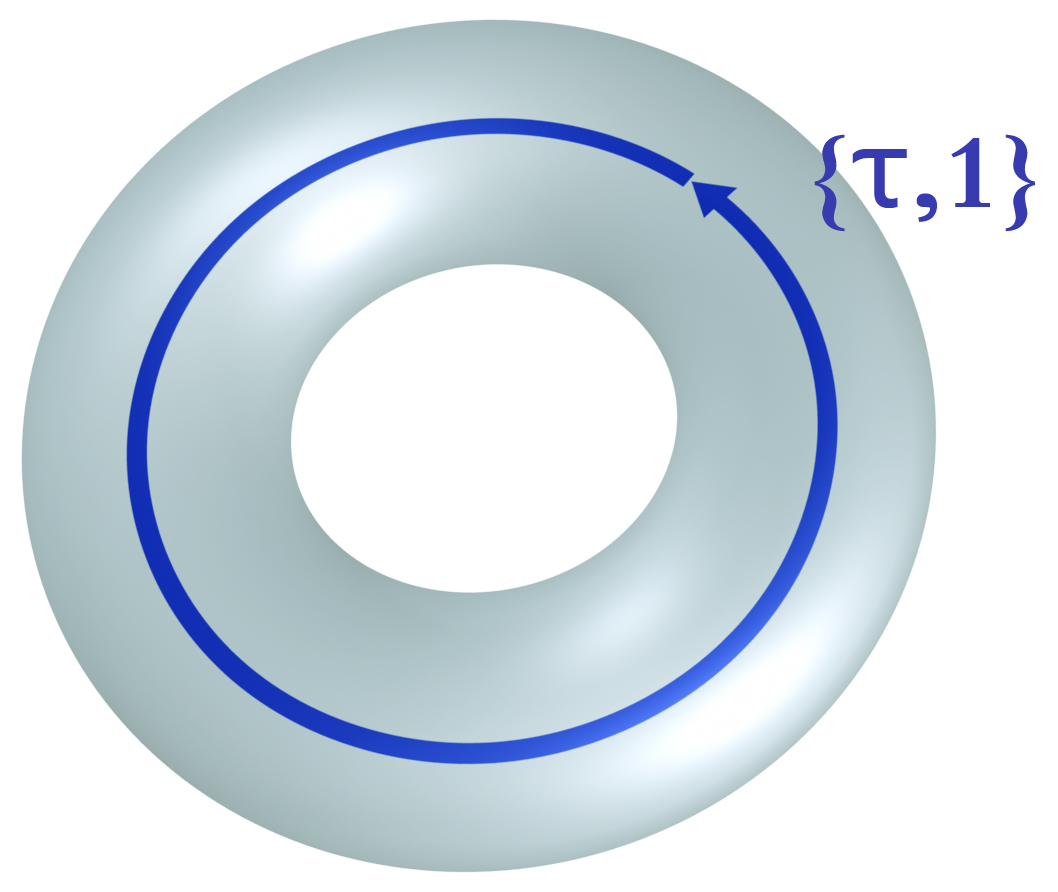}
\label{fig:fiboTorus}}
\caption{(a) \FF on a cylinder with both boundaries characterized by the  Lagrangian subset $\{1,\tau\bar\tau\}$. (b) Single Fibonacci phase on a torus. The two systems are equivalent.}
\end{figure}
Similarly, we can consider placing the \FF\ on a surface with three holes, or the ``pants diagram'' (a special case of Fig. \ref{fig:multiHole}). The three boundaries must again be characterized by $L_{\tau\bar\tau}$. On this surface, when two confined anyon loops $a$ and $b$ respectively winding two of the three cycles (holes) merge to an anyon loop $c$ winding the third cycle, the three loops of anyons must admit a fusion channel. The GSD counting in this scenario then boils down to the formula:
\be\label{eq:FiboGSDpants}
GSD_{\text \FF}^{\text{pants}}=\sum_{a,b,c\in\T_{L_{\tau\bar\tau}}}N^c_{ab}=5,
\ee
where  for a given $c$, the fusion matrix element $N^c_{ab}$ is the multiplicity of $c$ in the fusion product of $a$ and $b$. This again agrees with the expected result following from
\be
GSD_{\text \FF}^{\text{pants}}=GSD_{\text{Fibo}}^{\text{genus-2 torus}}.
\ee  
We note that such a correspondence between a ``doubled phase'' on a surface with gapped boundaries characterized by condensates of all the \emph{diagonal} pair---the analogues of $\tau\bar{\tau}$---and the undoubled phase on a closed surface is in fact generic. This correspondence offers a non-trivial check of our methods in large classes of non-Abelian phases expressible as a doubled phase using well-known results of GSD of phases on closed surfaces, and by which we find perfect agreement.

More generically, the boundaries could also be characterized by different $L_i$'s, which would contain anyons not mutually local. One would have to work out the condensed phase
$\T_{L_i} $ at each boundary, and then further reduce the number of conserved anyons, which correspond roughly to finding an intersection of the $\T_{L_i}$'s. The fusion between the remaining conserved anyons again determine the GSD. There is not to date a fully systematic procedure dealing with multiple sets of non-mutually local condensates, but we will exemplify how this is to work by a non-trivial example in the next section. 
In summary (Fig. \ref{fig:multiHoleGenus}), assuming that there are $M>3$ boundaries, respectively characterized by (potentially identical) Lagrangian subsets $L_i$'s,\be\label{eq:GSDgenBC}
GSD_{\{L_i\}}=\sum_{\{a_i,b_i\}}N_{a_1a_2}^{b_1}\prod_{i=1}^{M-4}N_{a_{i+2}b_{i}}^{b_{i+1}} N_{a_{M-1} b_{M-3}}^{\bar{a}_M},
\ee
where $\{\T_{L_i}\}:=\cap_{i=1}^M \T_{L_i}$, $b_i$'s are the intermediate fusion channels, and $\{a_i,b_i\}$ refers to summing over $\{\T_{L_i}\}$. Note that the product term exists only for $M\ge 5$.

\textbf{Counting GSD via charge transport.}
As alluded to at the beginning and also in the discussion of Abelian phases, we can alternatively count the GSD by considering charge transport across boundaries. This amounts to counting the fusion channels to the trivial sector, between anyons across Lagrangian subsets $L_i$'s characterizing the boundaries. Recall that \FF on a cylinder has a GSD given in (\ref{eq:fiboGSDcyl}). The same result can be 
obtained by counting the fusion channels between condensed anyons on the two boundaries to the trivial sector. Specifically, since there is only one trivial fusion channel in $\tau\bar\tau\x\tau\bar\tau=1+1\bar\tau+\tau\bar 1+\tau\bar\tau$, together with the obvious trivial fusion channel $1\x 1=1$, there are exactly 2 trivial fusion channels. Such an agreement between the two different ways of counting the GSD shouts for a generalization of the aforementioned LTW argument for FQHS to the case of non-Abelian topological phases. i.e. In a non-Abelian (gauge) theory,  there should exist some \emph{adiabatically changing} Wilson (anyon) loop, e.g., $\chi$ around the cylinder, which pumps a unit of the condensed  $\tau\bar\tau$ from one boundary to the other (Fig. \ref{fig:fiboCylinder}). 

To be seen in the next section, such a flux-charge correspondence would work only if each $L_i$ also includes the \emph{multiplicity} data of a condensed anyon---the number of condensed sectors contained in each anyon in $L_i$ that splits under condensation. This extra twist in the story makes GSD counting by confined anyons more natural.

\section{$\Z_2^3$ twisted quantum double}
We now present here a fascinating example---the $\Z_2^3$ twisted quantum double (TQD)\cite{Propitius1995,Hu2012a,Mesaros2011}---that bears more than one set of nontrivial gapped boundary conditions. As a twisted version of the $G=\Z_2^3$ Kitaev model, this model contains 22 distinct anyons. As in other gauge theories, the electric charges are representations of the gauge group $G$, and as such they come in 8 distinct types.  We can denote them by $E_{e_1e_2e_3}$, where $e_i \in \{0,1\}$, corresponding to the trivial and nontrivial one dimensional representations of each of the three $\Z_2$ groups in $G$. The rest of the anyons are  $14$ non-Abelian dyons all with quantum dimension $2$, denoted by $D^\pm_{m_1m_2m_3}$, where $m_i\in \{0,1\}$, and $m_1m_2m_3\neq 000$. Their properties, such as fusion rules, are detailed in the supplemental material. We focus on two distinct admissible Lagrangian subsets, $L_{\textrm{E}}$ and $L_{\textrm{D}}$. Set $L_{\textrm{E}}=\{E_{e_1e_2e_3}\}$ contains all the electric charges. As explained in the supplemental material, the condensed phase $\T_{\textrm{E}}$ contains the new trivial sector and $7$ non-trivial confined anyons that descend from the dyons, where $D^{\pm}_{m_1m_2m_3} \to 2 d_{m_1m_2m_3}$.
This gives, on a cylinder with both boundaries characterized by $L_\mathrm{E}$ 
\be 
 GSD^{L_{\textrm{E}}\, \textrm{boundaries on cylinder}}_{\Z_2^3 \,\textrm{TQD}} = 8.
\ee This immediately agrees with the result obtained by considering allowed charge transport across the boundaries, i.e. the number of fusion channels between the condensed anyons in the top and bottom boundary that fuse to the trivial sector. 
More interesting is the boundary characterized by $L_{\textrm{D}}$, where 
\be
L_{\textrm{D}} = \{2D^{+}_{100}, E_{0e_2e_3}\}.
\ee
The resultant condensed phase $\T_{\textrm{D}}$ contains again 8 distinct sectors, 7 of which confined and descended from the other dyons and electric charges of the form $E_{0e_2e_3}$. A very special thing arises here: The condensed anyon  $D^{+}_{100} \to 1+ 1$ splits into two copies of the vacuum in the condensed phase $\T_{\textrm{D}}$. This is unlike the examples   encountered above, where each sector appearing in the condensate only splits into the trivial sector once! To make that information explicit,  we have included the dyon $D^{+}_{100}$ twice in defining the set  $L_{\textrm{D}}$ above. Now the number of confined sectors in $\T_{\textrm{D}}$ would indicate that on a cylinder, 
\be GSD^{L_{\textrm{D}}\, \textrm{boundaries on cylinder}}_{\Z_2^3 \,\textrm{TQD}} = 8.
\ee
again. However, a naive count of allowed charge transport across the boundaries gives only 5 channels, if we count $D^+_{100}$ only once. The only way to recover a match between these two ways of counting is to take $D^+_{100}$ literally as appearing twice, so that they alone contribute $2^2= 4$ fusion channels between the top and bottom boundaries to the trivial sector, instead of only one as in the naive count. Then we recover a GSD =$ 4+ 4 = 8$.  
{\bf We therefore \emph{postulate} that the generalization of the Lagrangian subset in non-Abelian phases must include specifying the \emph{multiplicity} of a condensed anyon---the number of condensates that is actually contained in the splitting of the anyon after anyon condensation.} We have tested this postulate in this model in surfaces with more boundaries and found that the counting via charge-transport across boundaries continue to match the analysis via confined sectors in the condensed phase. We have also checked our postulate in the quantum double model with group $G=D_3$. The GSD due to a condensate involving multiplicities greater than 1 again supports our postulate.

\begin{figure}[ht]
\includegraphics[width=0.28\textwidth]{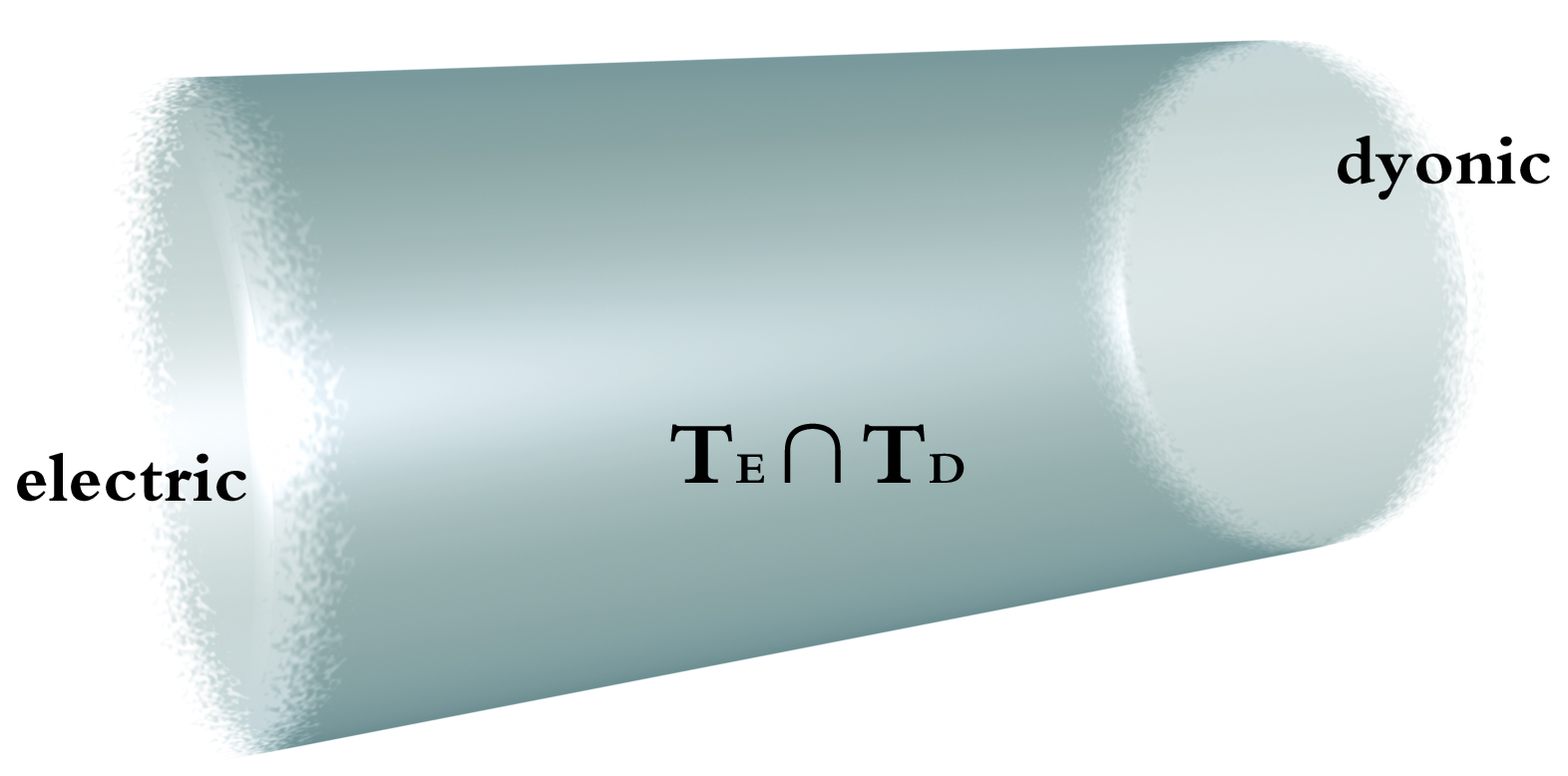}
\caption{$\Z_2^3$ twisted quantum double on a cylinder with the left boundary characterized by electric condensation $L_{\textrm{E}}$ and the right one by dyonic condensation $L_{\textrm{D}}$. $GSD=|T_{\textrm{E}}\cap T_{\textrm{D}}|$.}
\label{fig:tqdSepcialCyl}
\end{figure}

To end the section, we return to the $\Z_2^3$ TQD model on a cylinder, with, now, the left boundary characterized by $L_{\textrm{E}}$ and the right by $L_{\textrm{D}}$. The two sets of anyons have exactly 4 fusion channels that can fuse to one, namely the fusion of the four shared electric charges. So the charge transport reasoning leads to the  interesting result:
\be
GSD^{L_{\textrm{E}}\, \textrm{left, $L_{\textrm{D}}$ right}}_{\Z_2^3 \,\textrm{TQD}} = 4.
\ee  The same result can be obtained via counting the confined sectors, as depicted in Fig. \ref{fig:tqdSepcialCyl}.
As aforementioned and detailed in the supplemental material , $\T_{\textrm{D}}$ has 8 sectors, conveniently denoted by
\be
\{1, d^{1,2}_{m_2m_3}, d|m_2m_3\neq 00\}, 
\ee
satisfying the following important fusion rule 
\be
d^{1(2)}_{m_2m_3} \otimes d = d^{2(1)}_{m_2m_3}.
\ee
Since $d$ descends from $E_{1 e_1e_2}$, it is is no longer conserved in the other boundary where all electric charges are in $L_{\textrm{E}}$ and condense. Hence, $d^{\pm}_{m_2m_3}$ becomes indistinguishable, and the GSD is determined by the following four states each with an anyon line winding the non-trivial cycle
\be
\{| 1\rangle, \,  |d_{01}\rangle,\,  |d_{10}\rangle,\,  |d_{11}\rangle\},
\ee 
leading again to a precise match. This is the first example to date of a non-Abelian phase whose GSD on an open surface with multiple boundary conditions is computed. 

{\it Conclusion:} We have achieved the long sought goal to count the GSD of a generic non-chiral non-Abelian topological order with boundaries, making use of insights and techniques developed in the past\cite{Wen1990b,Bais2002,Bais2009,Bais2009a,Kitaev2012,Wang2012,Hung2013,Gu2014a}. We note that very much analogous to prior analysis of gapped boundary conditions via the introduction of explicit boundary gapping terms, one could imagine that such an analysis is also possible for non-Abelian phases. Some preliminary work has been done for example in \cite{Cappelli2014}, in which explicit terms can be written down, whenever the bulk non-Abelian phase adopts a simple construction based on orbifolding Abelian ones. It would be useful to generalize these studies to other non-Abelian phases. Given the importance of robust GSD as a resource in TQC, our new understanding will be crucial towards finding experimental realizations and applications of topological orders.

\begin{acknowledgements}
We thank Juven Wang for explaining his work on Abelian phases to us, and William Witczak-Krempa for his comments on the manuscript. YW appreciates his mentor, Xiao-gang Wen, for his constant support and inspiring conversations. YW is grateful to Yong-Shi Wu for his hospitality at Fudan University, where the work was done, and for his insightful comments on the manuscript. YW also thanks his cousin, Wanxing Wang, who helped drawing the figures. LYH is supported by the Croucher Fellowship. Research at Perimeter Institute is supported by the Government of Canada through Industry Canada and by the Province of Ontario through the Ministry of Economic Development \& Innovation.

\textit{Note added}.---Recently we became aware of a related work by Lan, Wang, and Wen\cite{Lan2014}.
\end{acknowledgements}

\begin{appendix}
\section{The doubled Fibonacci Phase}\label{sec:Fibo}
We review the topological data of the doubled Fibonacci Phase here. 

Consider the doubled Fibonacci system, \FF, which has four anyons $\{1,\tau\bar\tau,1\bar\tau,\tau\bar 1\}$ of quantum dimensions $d_1=1,d_{\tau\bar\tau}=1+d_\tau$, and $d_{1\bar\tau}=d_{\tau\bar 1}=d_\tau$, with $d_\tau=\tfrac{1+\sqrt{5}}{2}$. Their self statistics are  $\theta_1=1, \theta_{\tau\bar\tau}=1, \theta_{1\bar\tau}=\exp(-2\pi i/10)= \overline{\theta_{\tau\bar 1}}$. Since $\theta_{\tau\bar\tau}=1$, $\tau\bar\tau$ is the only nontrivial anyon that may condense. Thus, the only admissible non-Abelian Lagrangian subset would be $L_{\tau\bar\tau}=\{1,\tau\bar\tau\}$. Moreover, because $d_{\tau\bar\tau}>1$, $\tau\bar\tau$ must split in order to condense\cite{Bais2002,Bais2009,Bais2009a,Hung2013,Gu2014a,Bais2014}: $\tau\bar\tau=1+\chi$, where the trivial part $1$ is the actual condensate, and  $\chi$ with $d_\chi=d_\tau$ is some topological sector in the phase $\T_{L_{\tau\bar\tau}}$ after the condensation of $L_{\tau\bar\tau}$. It is easy to check that in $\T_{L_{\tau\bar\tau}}$ all nontrivial anyons are confined and that $\chi$, $1\bar\tau$, and $\tau\bar 1$ are indistinguishable. Namely, $\T_{L_{\tau\bar\tau}}=\{1,\chi\}$ with $\chi\x\chi=1+\chi$, behaving just like the Fibonacci anyon $\tau$.

\section{The $\Z_2^3$ Twisted Quantum Double}
We review the details of the $\Z_2^3$ TQD model here, which is characterized by the $3$-cocycle $\omega \in H^3(\Z_2\times\Z_2\times\Z_2,U(1))$, 
\be
\omega(x,y,z) = e^{\ii\pi x_1 y_2 z_3},
\ee
where each of $x,y,z$ is a three component vector, and each of its component takes values in $\{0,1\}$.

The anyons in this theory fall into two categories: pure charges and dyons. Each pure charge is labelled by a three component vector $E=(e_1,e_2,e_3)$,  where $e_i\in \{0,1\}$.
The dyons are each two dimensional irreducible projective presentation of $\Z_2\times\Z_2\times\Z_2$. A dyon is labeled by a $3$-vector $M=(m_1,m_2,m_3) \in \Z_2\times\Z_2\times\Z_2$, and for each $M$ there are two corresponding distinct representations, labeled $\pm$, which can be taken as the charge label.  The table summarises the distinct topological sectors in the theory and their self-statistics\cite{Propitius1995}.

\begin{table}[ht!]
\centering
 \begin{tabular}{ |c||c|c| }
 \hline
Top. Sector & Self-statistics & q.d.= $d_a$ \\
\hline
$E_{E=(e_1,e_2,e_3)} $ & 1 & 1\\
\hline
$D^{\pm}_{M}\vert_{ [M=(m_1,m_2,m_3),\,\, {M\neq (0,0,0)}]}$ & $\pm 1$ & 2\\
\hline
$D^{\pm}_{M}\vert_{[M=(1,1,1)] }$ & $ \pm i$ & 2\\
\hline
 \end{tabular}
\caption{
 Different topological sectors of $D^\omega(\Z_2\times\Z_2\times\Z_2)$
}
\label{tobsec}
\end{table}

We will explore two boundary conditions characterized by two different generalized Lagrangian subsets and work out the corresponding condensed phase for each of them. We shall not need to make use of the entire fusion algebra of the anyons of this model, for which we refer the reader to Eqs. (2.5.28) and (2.8.14) through (2.8.17) in \cite{Propitius1995}. The relevant fusion rules will be shown explicitly where they are used in the subsequent discussion.  
\subsection{Electric condensate}

This is the simplest scenario, in which 
\be
L_{\textrm{E}} = \{E_{e_1e_2e_3}\},
\ee
meaning that the boundary is characterized by the simultaneous condensation of all the electric charges,
all having trivial self statistics and mutual statistics among themselves.

All members in $L_{\textrm{E}}$ have quantum dimension unity, so they do not split any further in the condensed phase $\T_{\textrm{E}}$. The dyonic anyons have quantum dimension $2$, and might potentially split into two pieces.

Let us assume that 
\be
D^{\pm}_{M} = d^{\pm,1}_{M} \oplus d^{\pm,2}_{M}  .
\ee

Now using the fusion rules
\be
D^{\pm}_{M} \otimes E_{e_1e_2e_3} = D^{\pm (-1)^{e_1+e_2 + e_3+1}}_{M}.
\ee

Replacing these anyons by their decomposition in $\T_{\textrm{E}}$ immediately implies that 
$d^{+,i}_M= d^{-,i}_M \equiv d^i_M$.
Now recall that $D^{+}_M$ and $D^-_M$ have different topological spins  (Table \ref{tobsec}), implying that $d^i_M$, descended from both sectors, have ill-defined statistics, and thus must be confined in the condensate.  This is a proof that the choice of the set of condensate $L_{\textrm{E}}$ leads to complete confinement of the phase, and thus a gapped boundary. Now,
\be
D^{\pm}_{m_1m_2m_3}\otimes D^{\pm}_{m_1m_2m_3} = E_{E_1} + E_{E_2} + E_{E_3} + E_{E_4},
\ee
where the precise value of $E_i$ depends on the vectors $\{m_1,m_2,m_3\}$. What is important however is that as soon as we replace it by their decompositions, one concludes that
\be
d^i_M \otimes d^i_M = 1 = d^1_M \otimes d^2_M.
\ee
For a unitary theory in which the conjugate of an anyon is unique, this implies that
\be
d^1_M = d^2_M = d_M.
\ee
We therefore end up with exactly 8 distinct sectors in $\T_{\textrm{E}}$.

\subsection{Dyonic condensate}

Now we move on to the more interesting gapped boundary condition characterized by a dyonic condensate $L_{\textrm{D}}$.

To begin with we would like to pick $D^{+}_{100}$ as a condensate.
This has to split into two parts:
\be
D^{+}_{100} \to 1 \oplus c.
\ee
Taking into account that other dyons also potentially split into two pieces, we write 
\be D^{\pm}_{M} \to c^{M,\pm}_1 \oplus c^{M,\pm}_2.\ee
As we will conclude below, $D^{\pm}_M$ should split if $\T_{\textrm{D}}$ remains unitary, such that each anyon and its conjugate only fuse to the trivial sector once. 

Using the fusion relations
\begin{eqnarray}
D^+_{100}\otimes E_{0e_2e_3} &=&  D^+_{100} ,\\
D^+_{100}\otimes D^+_{100} &=& \sum_{e_2,e_3} E_{0e_2e_3},
\end{eqnarray}
one can conclude that 
\be
E_{0e_2e_3} \to 1, \,\, c =1.
\ee

Then the following fusion rules
\begin{eqnarray}
D^+_{100}\otimes D^-_{100} &=& \sum_{e_2,e_3} E_{1e_2e_3}, \\
D^+_{100}\otimes E_{1e_2e_3} &=&  D^-_{100} , \\
D^-_{100}\otimes D^-_{100} &=& \sum_{e_2,e_3} E_{0e_2e_3},
\end{eqnarray}
immediately imply that
\begin{eqnarray}
E_{1e_2e_3} &\to& d, \\
D^-_{100} &\to& c^{100,-}_1 \oplus c^{100,-}_2= 2d,\\
d\otimes d &=&1,
\end{eqnarray}
the last equality following also from  $E_{1e_2e_3}^2 =1$.
This immediately suggests that $E_{1e_2e_3}$ and $D^-_{100}$ are confined because they have different topological spins and yet decompose to the same anyon.
Next we have
\begin{eqnarray}
D^{\pm}_{0m_2m_3} &\otimes& E_{0e_2e_3} = D^{ \mp(-)^{\gamma_1+ \gamma_2}  }_{0m_2m_3}, \\
\gamma_i &=&  \frac{(2m_i-1) (2e_i-1) +1}{2}.
\end{eqnarray}
This implies that 
\begin{eqnarray}
&&c^{M,+}_1 \oplus c^{M,+}_2 =  c^{M,-}_1 \oplus c^{M,- }_2, \,\,\, \\
&&M\in \{ (1 (0),m_2,m_3)\}\vert_{\textrm{excluding} \,\,m_2=m_3=0}.
\end{eqnarray}
Therefore, we are identifying 
\be
c^{M,+}_i =  c^{M,-}_i,
\ee
and again these anyons become confined. This implies that all non-trivial sectors are
confined in the condensate, and 
\be
L_{\textrm{D}} = \{2 D^{+}_{100}, E_{0e_2e_3}\}
\ee
satisfies the desired condition of a generalized Lagrangian subset.

Now
\be
D^+_{100} \otimes D^{\pm}_{0m_2m_3} = D^{+}_{1m_2m_3} \oplus D^{-}_{1m_2m_3},
\ee
leading to identifying 
\be
c^{0m_2m_3,\pm}_i = c^{1m_2m_3,\pm}_i \equiv d^{i}_{m_2m_3},\,\, (\textrm{excluding} \, m_2=m_3=0).
\ee
where we further simplify notations by introducing $d^{i}_{m_2m_3}$.
Using 
\begin{eqnarray}
&&D^{+}_{0m_2m_3} \otimes E_{1e_2e_3} = D^{+ (-1)^{\gamma_1+\gamma_2}}_{0m_2m_3}, \\
&&D^{+}_{0m_2m_3}  \otimes D^{+}_{0m_2m_3} \nonumber \\
&&=\sum_{e_1} E_{e_100} + E_{e_1m_3m_2} ,
\end{eqnarray}
one concludes that 
\begin{eqnarray}
&&(d^i_{m_2m_3})^2 =1, \,\, d^1_{m_2m_3} \otimes d^2_{m_2m_3} = d, \,\, \\
&&d^1_{m_2m_3} \otimes d = d^2_{m_2m_3},  \\
&&d^i_{m_2m_3} \otimes d^i_{m'_2m'_4} = d^1_{(m_2+ m'_2)( m_3 +m'_3)}, \\
&&d^1_{m_2m_3} \otimes d^2_{m'_2m'_3} = d^2_{(m_2+ m'_2)( m_3 +m'_3)},
\end{eqnarray}
where the addition appearing in the subscripts are defined only modulo 2.
\end{appendix} 

\bibliographystyle{apsrev}
\bibliography{StringNet}
\end{document}